# On-chip single photon emission from an integrated semiconductor quantum dot into a photonic crystal waveguide

Andre Schwagmann[1,2], Sokratis Kalliakos[1], Ian Farrer[2], Jonathan P. Griffiths[2], Geb A. C. Jones[2], David A. Ritchie[2], and Andrew J. Shields[1,a]

[1] *Cambridge Research Laboratory, Toshiba Research Europe Limited, 208 Science Park, Milton Road, Cambridge, Cambridgeshire CB4 0GZ, United Kingdom*

[2] *Cavendish Laboratory, University of Cambridge, Madingley Road, Cambridge, Cambridgeshire CB3 0HE, United Kingdom.*

We demonstrate the in-plane emission of highly-polarized single photons from an InAs quantum dot embedded into a photonic crystal waveguide. The spontaneous emission rates are Purcell-enhanced by the coupling of the quantum dot to a slow-light mode of the waveguide. Photon-correlation measurements confirm the sub-Poissonian statistics of the in-plane emission. Under optical pulse excitation, single photon emission rates of up to 19 MHz into the guided mode are demonstrated, which corresponds to a device efficiency of 24%. These results herald the monolithic integration of sources in photonic quantum circuits.

[a] Electronic mail: andrew.shields@crl.toshiba.co.uk

Due to their immunity to decoherence, single photons (SP) are excellent candidates for the transfer of quantum information in the emerging fields of quantum information processing and quantum communications.[1-4] Logic gates for linear optical quantum computing that involved a few photons have already been demonstrated with discrete SP sources based on laser pumping of non-linear crystals.[5,6] However, scaling to large photon numbers will require the development of on-demand photon sources that can be integrated into a quantum photonic circuit.

Semiconductor quantum dots (QDs) allow for this on-chip integration, and proof-of-principle experiments with the emission directed out of the plane have shown that the excitonic recombination in QDs can be an on-demand SP source both for optical and electrical carrier injection.[7-11] High efficiencies and short excitonic lifetimes are achievable by placing the QD into a microcavity, where the spontaneous emission (SE) rate can be enhanced by the Purcell effect.[8,12,13] Embedding QDs into photonic crystal (PC) structures that allow for confinement and manipulation of light in two or three dimensions is very attractive for their integration into quantum photonic circuits. Two-dimensional PC structures are generally comprised of an array of etched holes in a suspended slab, and light can be guided in-plane by PC waveguides (PCWGs) formed by the omission of one or more lines of holes. An emitter coupled to a propagating mode in PCWGs can also be subject to Purcell enhancement due to slow-light effects at the edges of the Brillouin zone[14-18] leading to a high extraction efficiency and broadband operation. However, direct evidence of in-plane SP emission from an integrated SP source has not been reported yet. Here, we exploit the guiding properties of PCWGs and demonstrate a device with a high in-plane SP emission rate into the propagating mode and 24% efficiency under pulsed optical excitation. The potential of such a device as an on-chip SP source has been highlighted in earlier studies,[19,20] and our device is conceptually based on a recent proposal[14] with two important modifications that appeared to be crucial for this experimental demonstration. First, in our devices, SPs are transmitted by the higher-order propagating mode instead of the fundamental mode, thus requiring structures with larger lattice constants that are less susceptible to fabrication imperfections.[21] Second, we optimized the hole radius to lattice constant ratio r/a for our structures by finite-difference time-domain (FDTD) calculations in order to achieve higher theoretical Q-factors.

We fabricated a unidirectional PC W1 slab waveguide by removing one line of air holes along the Γ-K direction in a hexagonal lattice PC patterned in GaAs by wet and dry etching.[22] A layer of low-density InAs QDs that serve as quantum emitters is embedded at the center of the slab. A scanning electron microscope image of such a device is shown in Fig. 1a. We probe the directed light emission into the PCWG by optically exciting with laser pulses from the top of the slab while collecting from the PCWG exit, as illustrated in Fig. 1b.[22]

Figure 2a shows the calculated photonic band structure of the corresponding infinite waveguide, obtained by the plane-wave expansion method for TE-like modes (electric field along the slab plane). There are two distinct guided y-polarized modes with almost-zero dispersion close to the Brillouin zone edge (black lines) within the photonic bandgap and below the light line. These slow-light enhanced modes appear as sharp resonances in the mode structure obtained by FDTD calculations for our unidirectional device, as shown in Fig. 2b. The energies of the two y-polarized sharp modes M0 and M1 at 0.326 eV•μm (1.038 eV for a = 314 nm) and 0.438 eV•μm (1.398 eV) are in agreement with the guided modes from (a) and the small shifts are attributed to the transition from an infinite to the finite structure, similar to those observed in ref. 14 but with opposite sign, probably due to design differences. The low-intensity broad peaks are due to Fabry-Perot resonances between the PC mirror and the GaAs-air interface. We focus on the high-energy mode (M1) for slow-light enhancement of the SE and SP transmission. The calculated Q-factor is 7000 and its electric field profile $E_y$ is shown in Fig. 2c.

Figure 1c shows the in-plane photoluminescence spectrum, which is dominated by two intense sharp lines. We identify these lines as the exciton (X) and biexciton (XX) transitions by their approximately linear ($\sim P^{0.8}$) and quadratic ($\sim P^{1.8}$) power dependencies, respectively (see Fig. 1d). Lorentzian fitting reveals the presence of a slow-light PCWG mode (broad red curve, Fig. 1c). Its appearance at 1.386 eV is in good agreement with the theoretically predicted value (at 1.398 eV) and allows the identification as the target mode discussed in the previous paragraph. The measured Q-factor of ~ 600 is approximately one order of magnitude smaller than predicted, presumably due to fabrication imperfections. However, this allows both the exciton and the biexciton line to be in resonance with the mode.

The radiative lifetimes of the QD populations are shown in Fig. 3b. For the X (black discs) and XX (red discs) lines we measure lifetimes of ~ 0.85 ns and ~ 0.57 ns, respectively. The excitonic recombination time is decreased compared to QDs in unprocessed portions of the sample (bulk QDs), where we measure an average lifetime of ~ 1.45 ns (blue discs, Fig 3b). From this we infer a moderate Purcell factor of 1.7 for the exciton line. Stronger Purcell enhancement can be achieved by a better spectral overlap with the mode via temperature tuning.[22] On the other hand, for an off-resonant emitter (at 1.397 eV, Fig. 1c) we measure a lifetime of ~ 2.7 ns (purple discs), which is a manifestation of Purcell suppression. We attribute the relatively low Purcell factors to poor spatial overlap of the emitter with the maximum of the mode's electric field. In addition, we confirm light injection from the emitter into the guided mode M1 by analyzing the polarization of the detected photons (as shown in Fig. 3a). We find the emitted light to be strongly polarized (90%) along the y-direction, which matches the polarization of the propagating mode. The weak emission that is polarized along the vertical (z) direction is attributed to polarization scattering.[17]

Strong photon antibunching is a characteristic of SP sources, and it can be measured using a Hanbury-Brown and Twiss setup. The recorded coincidences give a direct measure of the second-order correlation function $g^{(2)}(\tau)$: $g^{(2)}(\tau) = \langle I(t)I(t+\tau) \rangle / \langle I(t) \rangle \langle I(t) \rangle$, where $I(t)$ is the photon intensity at time $t$. Figure 3c shows a histogram of the coincidence events measured for the exciton line under pulsed excitation. The suppressed peak at $\tau = 0$ is a clear signature of SP emission while the negative correlation in the adjacent peaks is attributed to memory effects in the QD.[23] Analysis of the histogram allows us to deduce a value of $g^{(2)}(0) = 0.23$. This suggests that, in our device, the rate of multiphoton emission is less than a quarter of that of an attenuated laser of the same average intensity. The remaining multiphoton events at $\tau = 0$ are attributed to stray light from the buffer and substrate layers.

We further investigate our device's performance as in-plane SP source by measuring the SP emission characteristics as a function of excitation power. Figure 3d shows both the values of $g^{(2)}(0)$ (red squares) and the corresponding count rate (black squares) versus excitation power. We find that even at saturation $g^{(2)}(0)$ is always lower than 0.5. The progressive increase of multi-photon emission is due to QD saturation and emission from other QDs at high pump rates.[24] At the same time the detected rates approach 120 kHz. After multi-photon emission event correction,[22] we find a detected SP emission rate of 100 kHz (black open circles, Fig. 3d). Given that the detection efficiency of our setup is estimated to be 1% (see ref. 22) we determine an emission rate into the first lens of 10 MHz. For a unidirectional PCWG device like ours, the photon extraction efficiency has been calculated to be 53%.[14] This allows us to estimate an internal SP emission rate into the guided mode of the PCWG of 19 MHz. For our pump repetition rate of 80 MHz, this translates to an internal device efficiency of 24%.

In conclusion, we have fabricated an optimized unidirectional PCWG with an embedded QD as quantum emitter and demonstrated the on-chip injection of single photons into the guided mode with an efficiency of 24%. The Purcell enhancement of the SE rate provides direct evidence of the coupling of the emitter to a slow-light mode of the waveguide. These results mark a cornerstone for the planar integration of the SP source with quantum photonic integrated circuits.

**Acknowledgements.** This work was partly supported by the EU through the Integrated Project Q-ESSENSE (contract no. FP7/2007–2013). A. S. acknowledges financial support from EPSRC, Cambridge European Trust, Kurt Hahn Trust and the Thomas-Gessmann Stiftung. We thank D. J. P. Ellis and K. Cooper for help with the device processing and S. J. Dewhurst, D. Granados and A. J. Bennett for discussions.

# References

[1] E. Knill, R. Laflamme and G. J. Milburn, *Nature* **409,** 46-52 (2001).

[2] D. Bouwmeester, J-W. Pan, K. Mattle, M. Eibl, H. Weinfurter and A. Zeilinger, *Nature* **390,** 575-579 (1997).

[3] J. L. O'Brien, *Science* **318,** 1567-1570 (2007).

[4] A. J. Shields, *Nature Photon.* **1,** 215–223 (2007).

[5] A. Politi, M. J. Cryan, J. G. Rarity, S. Yu and J. L. O'Brien, *Science* **320,** 646-649 (2008).

[6] J. C. F. Matthews, A. Politi, A. Stefanov and J. L O'Brien, *Nature Photon.* **3,** 346 - 350 (2009).

[7] P. Michler, A. Kiraz, C. Becher, W. V. Schoenfeld, P. M. Petroff, L. Zhang, E. Hu and A. Imamoglu, *Science* **290,** 2282-2285 (2000).

[8] S. Strauf, N. G. Stoltz, M. T. Rakher, L. A. Coldren, P. M. Petroff and D. Bouwmeester, *Nature Photon.* **1,** 704-708 (2007).

[9] M. Pelton, C. Santori, J. Vučković, B. Zhang, G. S. Solomon, J. Plant and Y. Yamamoto, *Phys. Rev. Lett.* **89,** 233602 (2002).

[10] Z. Yuan, B. E. Kardynal, R. M. Stevenson, A. J. Shields, C. J. Lobo, K. Cooper, N. S. Beattie, D. A. Ritchie and M. Pepper, *Science* **295,** 102–105 (2002).

[11] W.-H. Chang, W.-Y. Chen, H.-S. Chang, T.-P. Hsieh, J.-L. Chyi and T.-M. Hsu, *Phys. Rev. Lett.* **96,** 117401 (2006).

[12] D. Englund, D. Fattal, E. Waks, G. S. Solomon, B. Zhang, T. Nakaoka, Y. Arakawa, Y. Yamamoto and J. Vučković, *Phys. Rev. Lett.* **95,** 013904 (2005).

[13] A. Badolato, K. Hennessy, M. Atatüre, J. Dreiser, E. L. Hu, P. M. Petroff and A. Imamoğlu, *Science* **308,** 1158-1161 (2005).

[14] V. S. C. Manga Rao and S. Hughes, *Phys. Rev. Lett.* **99,** 193901 (2007).

[15] T. Lund-Hansen, S. Stobbe, B. Julsgaard, H. Thyrrestrup, T. Sünner, M. Kamp, A. Forchel and P. Lodahl, *Phys. Rev. Lett.* **101,** 113903 (2008).

[16] E. Viasnoff-Schwoob, C. Weisbuch, H. Benisty, S. Olivier, S. Varoutsis, I. Robert-Philip, R. Houdré and C. J. M. Smith, *Phys. Rev. Lett.* **95,** 183901 (2005).

[17] S. J. Dewhurst, D. Granados, D. J. P. Ellis, A. J. Bennett, R. B. Patel, I. Farrer, D. Anderson, G. A. C. Jones, D. A. Ritchie and A. J. Shields, *Appl. Phys. Lett.* **96,** 031109 (2010).

[18] V. S. C. Manga Rao and S. Hughes, *Phys. Rev. B* **75,** 205437 (2007).

[19] S. Hughes, *Opt. Lett.* **29,** 2659 (2004).

[20] P. Yao, V. S. C. Manga Rao and S. Hughes, *Laser & Photon. Rev.* **4,** 499 (2010).

[21] S. Hughes, L. Ramunno, J. F. Young and J. E. Sipe, *Phys. Rev. Lett.* **94,** 033903 (2005).

[22] See supplementary material at [URL will be inserted by AIP] for sample preparation, optical characterization, detection efficiency and temperature tuning.

[23] C. Santori, D. Fattal, J. Vučković, G. S. Solomon, E. Waks and Y. Yamamoto, *Phys. Rev. B* **69,** 205324 (2004).

[24] S. Strauf, K. Hennessy, M. T. Rakher, Y.-S. Choi, A. Badolato, L. C. Andreani, E. L. Hu, P. M. Petroff and D. Bouwmeester, *Phys. Rev. Lett.* **96,** 127404 (2006).

**FIGURE CAPTIONS**

FIG. 1. (Color online) (a) Scanning electron microscope image of the suspended PCWG device similar to the one used in the experiments. The scalebar corresponds to 1 µm (b) Schematic illustration of the experiment's principle. The laser light (red cone) is focused onto the top of the PCWG slab to excite the QD. The generated single photons are guided along the waveguide and collected at its exit. (c) In-plane micro-photoluminescence spectrum of a single QD at $T$ = 5 K. Peaks indicated as X and XX correspond to excitonic and biexcitonic recombination. The broad red curve indicates the PCWG mode. (d) Integrated intensity of exciton (black discs) and biexciton (red discs) lines. The gray lines are fits $\sim P^{0.8}$ and $\sim P^{1.8}$.

FIG. 2. (Color online) (a) Calculated band structure inside the first Brillouin zone for the TE-like modes for an infinite PCWG with a hole radius to lattice constant ratio $r/a$ = 0.345. The black (gray) lines indicate y-polarized (x-polarized) modes. (b) Mode structure for a unidirectional PCWG slab with a y-polarized emitter. The vertical axis has the same units and scale as in (a). (c) Electric field profile Ey of the mode M1 along the slab plane.

FIG. 3. (Color online) (a) In-plane micro-photoluminescence spectrum of light polarized along y- (black line) and z- (red line) directions. (b) Lifetime measurements at $T$ = 5 K of the X (black discs) and XX (red discs) lines, of an off-resonant emitter (purple discs) and of a QD in the bulk with decay time close to the average extracted

from several QDs (blue discs). The gray lines are single exponential fits. (c) Photon coincidence histogram for the X line for light detected in-plane under pulsed excitation. (d) Detected SP count rates (black squares, left axis), SP rates after multiphoton correction (open circles, left axis) and second order correlation value $g^{(2)}(0)$ (red squares, right axis) as a function of excitation power. The shaded area indicates the operation point for the highest SP emission rate. The red dashed line shows the limit at $g^{(2)}(0) = 0.5$ for SP operation.

**FIGURE 1**

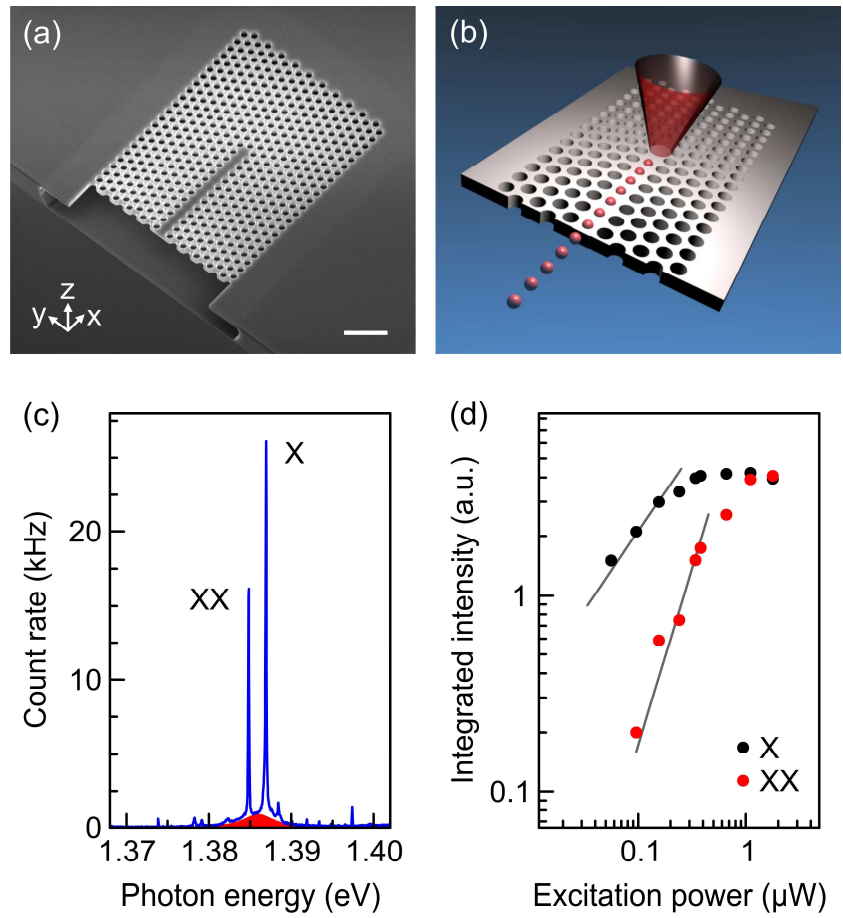

**FIGURE 2**

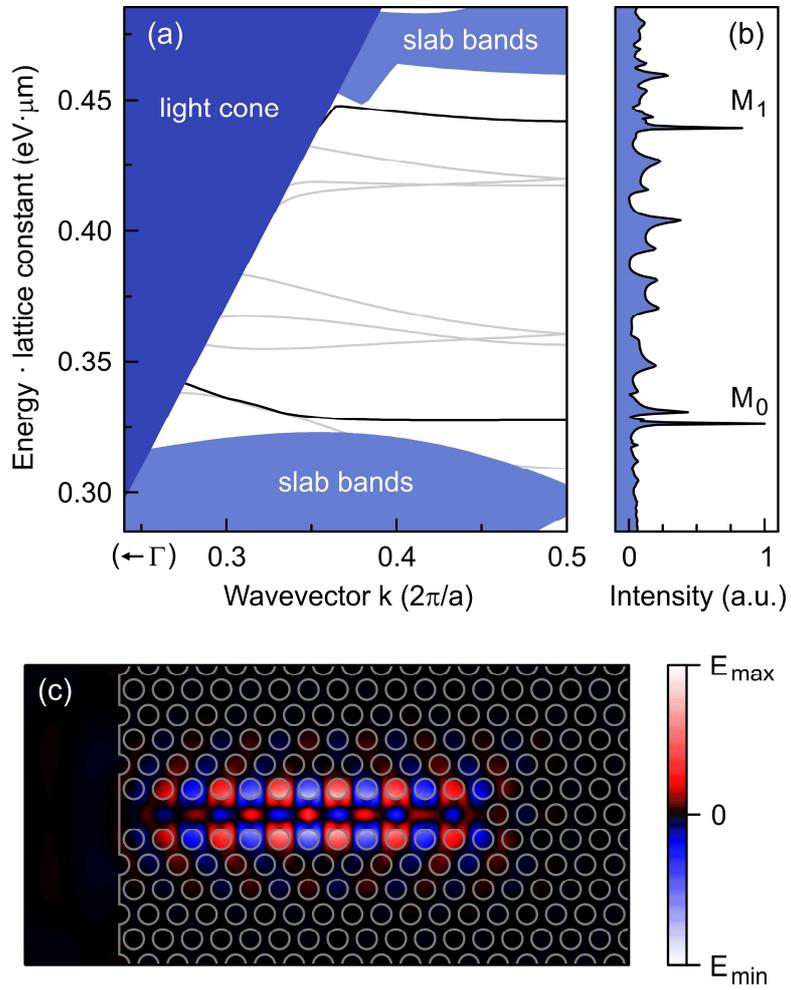

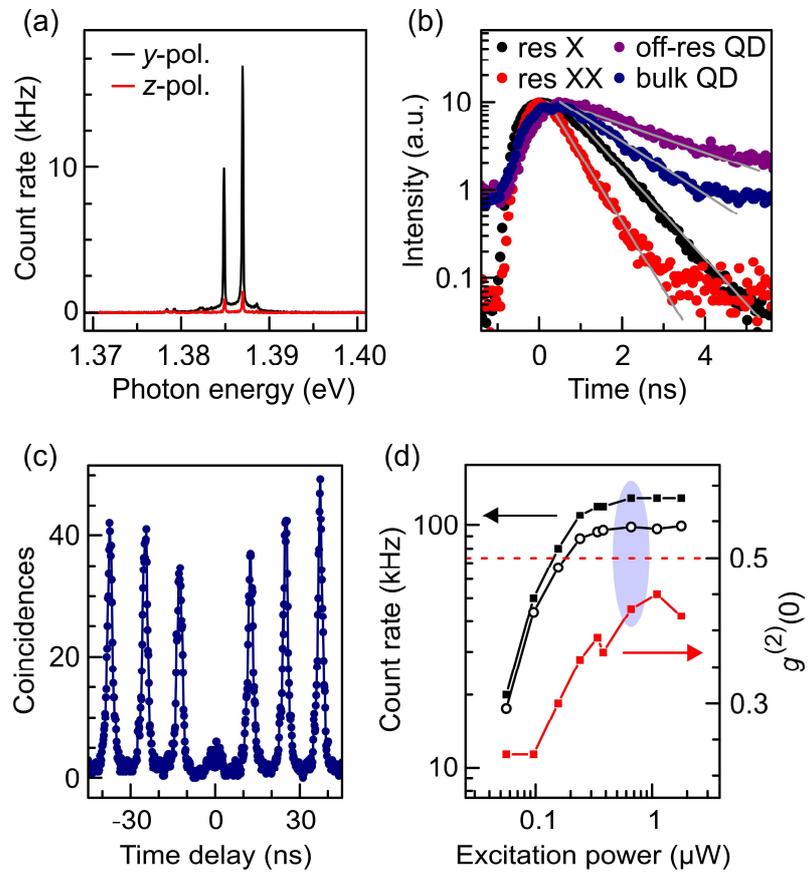